\documentclass[twocolumn,amssymb,fleqn]{revtex4} 
\usepackage{epsfig,amssymb,amsmath,graphicx,subfigure,hyperref}  
\usepackage{color}

\usepackage{multirow} 
\usepackage{tabularx}

\begin{document}
	
\title{Intrinsic statistical regularity of topological charges revealed in
		dynamical disk model}
	\author{Ranzhi Sun and Zhenwei Yao}
\email[]{zyao@sjtu.edu.cn}
\affiliation{School of Physics and Astronomy, and Institute of Natural
  Sciences, Shanghai Jiao Tong University, Shanghai 200240, China}
\begin{abstract}
    Identifying ordered structures hidden in the packings of particles is a
    common scientific question in multiple fields. In this work, we investigate
    the dynamical organizations of a large number of initially randomly packed
    repulsive particles confined on a disk under the Hamiltonian dynamics by
    the recently developed algorithm called the random batch method.  This
    algorithm is specifically designed for reducing the computational complexity
    of long-range interacting particle systems.  We highlight the revealed
    intrinsic statistical regularity of topological charges that is otherwise
    unattainable by the continuum analysis of particle density. We also identify
    distinct collective dynamics of the interacting particles under short- and
    long-range repulsive forces. This work shows the robustness and
    effectiveness of the concept of topological charge for characterizing the
    convoluted particle dynamics, and demonstrates the promising potential of
    the random batch method for exploring fundamental scientific questions
    arising in a variety of long-range interacting particle systems in soft
    matter physics and other relevant fields. 
\end{abstract}
	
	\maketitle

  \section{Introduction}

  The packing of interacting particles in confined geometries is a common theme
  in a host of soft matter
  systems~\cite{Lai1999,aastrom2000granular,donev2004unusually,weaire2008pursuit,mughal2011phyllotactic,manoharan2015colloidal,yao2019command}.
  In particular, the confluence of theoretical and experimental investigations in
  the past few decades on the static two-dimensional(2D) packing problem on both
  flat~\cite{Mughal2007,miguel2011laminar,Yao2013a,
  cerkaski2015thomson,soni2018emergent,silva2020formation,PhysRevE.104.034614}
  and
  curved~\cite{nelson2002defects,Bausch2003e,bowick2009two,mehta2016kinetic}
  spaces reveals the crucial role of topological defects in the organizations of
  the particles. Disclinations as a kind of fundamental topological defect
  emerge in the crystalline packings of particles on curved
  spaces~\cite{nelson2002defects,Bausch2003e,bowick2009two} or under
  mechanical~\cite{peach1950forces,hirth1982theory,miguel2011laminar} and
  thermal~\cite{halperin1978theory,strandburg1988two} stimuli.  In a triangular
  lattice, a $p$-fold disclination refers to a vertex of coordination number
  $p$, and it carries topological charge
  $6-p$~\cite{nelson2002defects,bowick2009two}.  According to the continuum
  elasticity theory, in analogy to electric charges, disclinations of the same
  sign repel and unlike signs attract~\cite{nelson2002defects}. These
  particle-like excitations are actively involved in several important physical
  processes, such as the screening of the substrate
  curvature~\cite{Bausch2003e,vitelli2006crystallography,azadi2014emergent} and
  the healing of the disrupted crystalline
  order~\cite{bowick2007interstitial,irvine2012fractionalization,yao2020fraction}.

  In our previous work on the inhomogeneous static packings of particles in
  mechanical equilibrium confined on a disk, the unique perspective of a
  topological defect allows us to reveal the characteristically negative sign of
  the total topological charge $Q$ and the associated hyperbolic
  geometry~\cite{Yao2013a}. Recently, the static disk model was extended to the
  dynamical regime~\cite{yao2021fast}. It was found that in the statistical
  sense, the hyperbolic geometry as reflected by the negative sign of the
  time-averaged total topological charge $\langle Q\rangle$ is preserved in the
  convoluted dynamical evolution of the particle
  configuration~\cite{yao2021fast}.

  However, in our previous work the dynamical disk model consisted of a relatively
  small number of particles as limited by the long computational
  time~\cite{yao2021fast}. It is thus still not clear if the revealed
  statistical regularity of topological charges is caused by the boundary
  effect or due to the intrinsic effect of the long-range force. Elucidating
  this issue yields insights into the intrinsic order in seemingly irregular and
  temporally-varying packings of particles arising in the physical systems of
  charged colloids~\cite{bowick2009two}, complex (dusty)
  plasmas~\cite{thomas1994plasma,morfill2009complex}, and a variety of charged
  entities at the nanoscale in electrolyte environments~\cite{Walker2011}.
  Furthermore, the relatively small number of particles in the previously
  studied disk model hinders one from performing a coarse-graining procedure to
  reveal and analyze the underlying flow patterns created by the repulsive
  forces of varying range.

  To address these questions, it is crucial to extend the dynamic disk model to
  the large-$N$ regime, where $N$ is the number of particles. The challenge for
  simulating the dynamics of a long-range interacting system comes from the high
  computational complexity: the computation time is on the order of $O(N^2)$
  per time step, and the whole simulation process involves up to one million
  simulation steps.  To solve this problem, we employ the recently developed
  algorithm called the random batch method (RBM), which is specifically designed for
  reducing the computational complexity of long-range interacting particle
  systems down to the order of
  $O(N)$~\cite{jin2020random,jin2022RBM_second_order}.  Numerical experiments
  show that the RBM is capable of capturing both transient dynamical
  structures~\cite{liang2022improved} and crucial statistical
  features~\cite{qi2023random}.

  The goal of this work is to employ the powerful computational tool of the RBM
  for exploring the statistical and dynamical physics of the disk model
  consisting of a large number of repulsive particles (up to 150,000).
  Specifically, we aim at identifying the intrinsic statistical regularity of
  topological charges, and studying the flow patterns under the repulsive forces
  of varying range. To this end, in our model the particles confined on the disk
  interact by the screened Coulomb potential; the range of the force is
  conveniently controlled by the screening length. In the initial state, both
  the positions and the velocities of the particles are randomly distributed. The
  motion of the particles under the interacting force is governed by the
  deterministic Hamiltonian dynamics. The equations of motion are numerically
  integrated by the standard Verlet method~\cite{frenkel2002understanding}.

  The main results of this work are presented below. By analyzing the
  trajectories of 150,000 particles, we show that under a long-range repulsive
  force, the time-averaged total topological charge $\langle Q\rangle$ is
  intrinsically negative corresponding to the hyperbolic geometry. Systematic
  simulations of the disk model at varying screening length show the regulation
  of the defect structure by the range of the repulsive force. These results
  suggest the robustness and effectiveness of the concept of topological charge
  for characterizing the convoluted particle dynamics at varying interaction
  range.  Furthermore, we reveal the distinct dynamical organizations of the
  particles under short- and long-range repulsive forces by analyzing the
  coarse-grained velocity fields, and we demonstrate the capability of the RBM in
  capturing featured dynamical structures such as vortices and radial flows. In
  this work, the revealed intrinsic statistical regularity of topological
  charges and the featured flow patterns advance our understanding on the
  convoluted dynamical organizations of geometrically confined repulsive
  particles.

	\section{Model and Method}
	
	
	The model consists of $N$ identical point particles of mass $m$ confined on a
	disk of radius $r_0$; the particles interact by the pairwise screened Coulomb
	potential. The evolution of the particle configuration is governed by
  the deterministic Hamiltonian dynamics (without introducing any thermal noise): 
	\begin{eqnarray}
		H=\sum_{i=1}^N\frac{{p_i}^2}{2m}+\sum_{i\neq
			j} \frac{\beta}{r_{ij}}\exp({-\frac{r_{ij}}{\lambda}})
	\end{eqnarray}
	where $\vec{p}_i$ is the momentum of particle $i$, $r_{ij}$ is the distance
	between particles $i$ and $j$, $\lambda$ is the screening length, and
	$\beta=q_0^2/(4\pi\epsilon)$, where $\epsilon$ is the dielectric constant. The
	rigid boundary condition is implemented by the reflection of particle
	trajectory at the boundary. In this work,
	the units of mass, length and time are the particle mass $m$, the mean
	distance of nearest particles $a$, and $\tau_0$. $\tau_0 = \sqrt{m
		a^3/\beta}$.  The units of velocity and force are thus $a/\tau_0$ and
	$\beta/a^2$. Under the assumption that the particles are arranged in a
	triangular lattice, the lattice spacing $a$ is estimated as 
	\begin{eqnarray}
		\frac{a}{r_0}=\sqrt{\frac{2\pi}{\sqrt{3}N}}. \label{ar}
	\end{eqnarray}

	In the initial state, the velocity $\vec{v}_{\mathrm{ini}}$ of each particle is random
	in both magnitude and direction; $v_{\mathrm{ini}}\in (0,1)$ and $\theta \in (0,2\pi)$,
	where $\theta$ is the angle of the velocity vector $\vec{v}_{\mathrm{ini}}$ with
	respect to some reference direction. The initial positions of the particles
	are randomly distributed by the procedure of random disk packing. The
	technical details are presented in Appendix A.  We employ the standard Verlet
	method to numerically integrate the equations of motion under the specified
	rigid boundary condition and the random initial
	conditions~\cite{frenkel2002understanding}. Specifically, the position of particle
	$i$ at time $t+2h$ is determined by its positions at times
	$t+h$ and $t$ via 
	\begin{eqnarray}
		\vec{x}_i(t+2h) = 2 \vec{x}_i(t+h) - \vec{x}_i(t) + \ddot{\vec{x}}_i(t+h)h^2  
		+ {\cal{O}} (h^4),\nonumber
	\end{eqnarray}
	where the step size $h=10^{-3}$ in simulations. 	 
	The most time-consuming procedure in the Verlet integration scheme for the
	long-range interacting particle system is the calculation of the forces. The
	computation time is on the order of $O(N^2)$ per time step. This imposes a
	challenge to explore the large-$N$ systems of interest.

	\begin{figure}[t] 
		\centering 
		\includegraphics[scale=0.3]{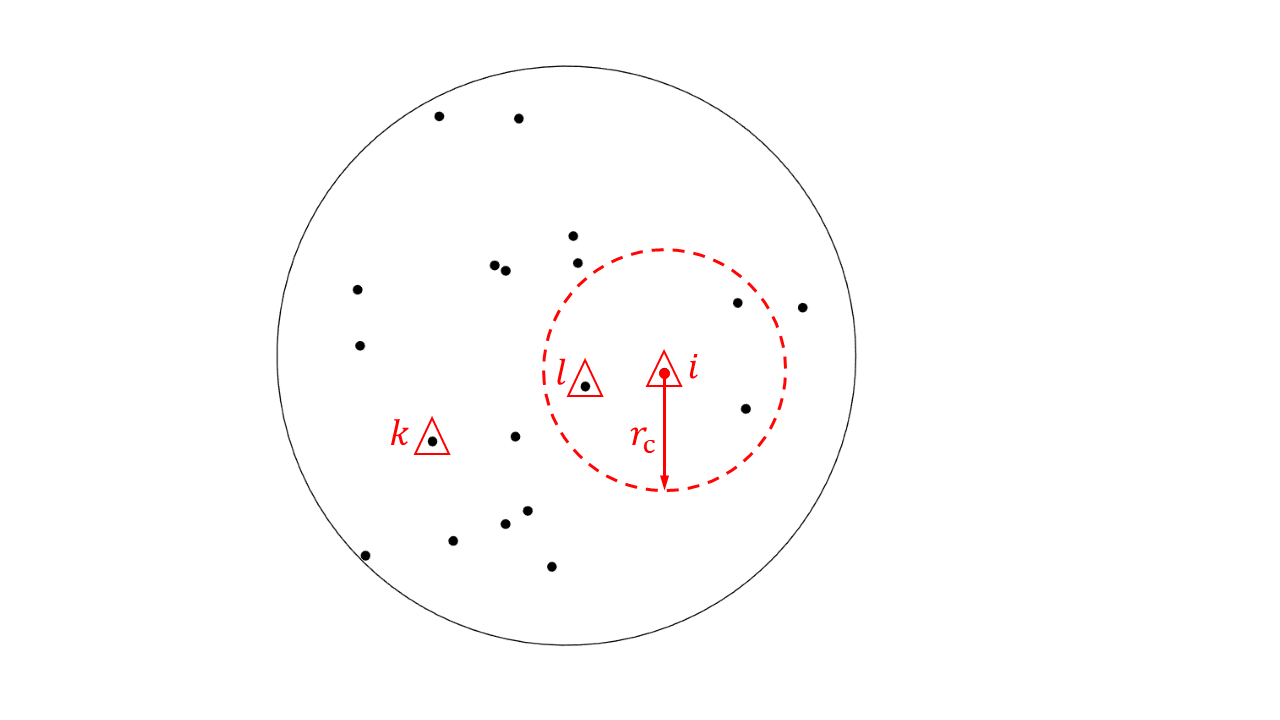}
		\caption{\label{fig:disk} Schematic plot for illustrating the random batch
			method. The particles on the disk are represented by the black dots. The particles
			marked by small triangles belong to the same batch $\Omega_p$. The adjacent particles
			surrounding the particle concerned (labeled $i$) are located in the dashed circle of
			radius $r_c$. } 
	\end{figure}

	
	To resolve this issue, we employ the recently developed algorithm named the
	random batch method (RBM) by
	mathematicians~\cite{jinRandomBatchMethods2020,jin2020random,jin2022RBM_second_order}.
	This algorithm is specifically designed for reducing the computing time for
	long-range interacting particle systems down to the order of $O(N)$.  In the
	following, we introduce our method of calculating force using RBM.	Consider a
	system consisting of a large number of point particles with pairwise
	interaction.  First, the interacting force on particle $i$ by particle
	$j$, $\vec{f}_{ij}$, is decomposed into two parts
	~\cite{jin2022RBM_second_order}:
	\begin{align} 
		\vec{f}_{ij}(r_{ij}) &= \vec{f}_{ij}(r_{ij}) [1-H(r_{ij} - r_c) ] \nonumber  \\
		& \ \ + \vec{f}_{ij}(r_{ij}) H(r_{ij} - r_c) \nonumber  \\
		&\equiv \vec{f}^{(1)}_{ij} + \vec{f}^{(2)}_{ij}	 \label{fij}
	\end{align}
	where the Heaviside step function
	\[ H(r_{ij} - r_c) = 
	\begin{cases}
		1 &   r_{ij} \geq r_c \\
		0 &   r_{ij} < r_c.
	\end{cases}\]
	$r_{ij}$ is the distance between particle $i$ and particle $j$.  $r_c$ is some
	cutoff distance that is comparable with the mean distance of nearest particles
	(see Fig.~\ref{fig:disk}). Now, the total force on particle $i$ can be written as
	\begin{align}
		\vec{F}_{i} &= \sum_{j\ne i} \vec{f}^{(1)}_{ij} + \sum_{j\ne i} \vec{f}^{(2)}_{ij},
		\label{Fi_latest}
	\end{align}
	The first term in Eq.~(\ref{Fi_latest}) is the sum of the forces exerted by all
	of the adjacent particles of particle $i$ within the circular region of radius
	$r_c$, which is calculated rigorously. The second term involves the forces
	from all of the remaining particles, and it is approximated by the RBM to
	promote computational efficiency.

	The schematic plot for illustrating the RBM is presented in
	Fig.~\ref{fig:disk}.  All of the $N$ particles on the disk are labeled from 1
	to $N$. For each instantaneous
    particle configuration in the time evolution, the $N$
	particles are randomly divided into a certain number of small batches by the
	particle labels (not by the particle positions). Each batch contains $p$
	particles. In Fig.~\ref{fig:disk}, the particles marked by small triangles
	belong to the same batch $\Omega_p$; $p=3$ in this case. Only the particles
	belonging to the same batch as particle $i$ are counted in the calculation for
	the second term in Eq.~(\ref{Fi_latest}). The summation of the forces from
	these particles is denoted as $\vec{F}_{\mathrm{batch}}$. Multiplying
	$\vec{F}_{\mathrm{batch}}$ by a factor of $(N-1)/(p-1)$ gives the approximate value for
	the second term in Eq.~(\ref{Fi_latest}). To conclude, the total force on
	particle $i$ is approximated by~\cite{jin2022RBM_second_order} 
	\begin{eqnarray} 
		\vec{F}_i = \sum_{r_{ij}< r_c} \vec{f}_{ij} + \frac{N-1}{p-1}\sum_{j\in\Omega_p}
		\vec{f}_{ij}(\vec{x}_{ij})H(|\vec{x}_{ij}| - r_c).	
		\label{Fi_rbm}
	\end{eqnarray}	
	In this work, the value of $p$ is taken to be $2$ by following the
	convention~\cite{jinRandomBatchMethods2020}.  Note that due to the random
	selection of the particles, the RBM brings in random forces on the particles,
	which results in a monotonous increase of the kinetic energy (temperature).
	For the sake of the conservation of total energy, we perform the procedure of
	cooling in simulations (see Appendix B).  Specifically, the velocity of each
	particle is rescaled by a common factor once the amount of the increased energy
	exceeds some threshold value, which is set to be about $5\%$ of the total energy
	in the initial state.

	Numerical experiments show that the RBM is capable of capturing both
	transient dynamical structures~\cite{liang2022improved} and crucial
	statistical features~\cite{qi2023random}. The RBM works well in a series of
	statistical and dynamical problems, such as the Dyson Brownian motion,
	Thomson's problem, stochastic dynamics of wealth and opinion
	dynamics~\cite{jinRandomBatchMethods2020}. This algorithm has wide
	applications in quantum
	physics~\cite{golse2021random,jin2020random,wang2021layer}, plasma
	physics~\cite{carrillo2022random}, hydromechanics~\cite{qi2023random}, machine
	learning~\cite{carrillo2021consensus}, chemistry and
	materials~\cite{liang2022improved}. In this work, the RBM is used as a
	powerful tool for exploring the physics of
	long-range interacting systems containing a large number of particles (up to
	150,000).

	\section{Results and discussion}
	
	In this section, we discuss the dynamics of the particles on
	the disk under both short- and long-range repulsive forces. In Sec. \uppercase\expandafter{\romannumeral3} A, the particle
	configurations are analyzed from the perspective of the underlying topological
	defect structure. Statistical regularity in the distribution of
	topological charge is revealed. 
	In Sec. \uppercase\expandafter{\romannumeral3} B, we analyze the characteristic coarse-grained velocity field, and
	we reveal the distinct flow patterns created by the short- and long-range repulsive forces. 
	We also discuss the relaxation of particle speed in both short- and long-range interacting systems.

	\subsection{Statistical regularity in the distribution of topological charge}

	We first briefly introduce the concept of a topological defect in a triangular
	lattice~\cite{Chaikin1995,nelson2002defects}. By the standard Delaunay triangulation, the
	neighbors of each particle on the plane are uniquely determined, and the
	topological charge $q_i$ for any particle $i$ is defined by the formula
	\begin{eqnarray}
		q_i = 6 - z_i,
	\end{eqnarray}
	where $z_i$ is the coordination number of particle
	$i$~\cite{nelson2002defects}. A particle of coordination number $z$ is called
	a $z$-fold disclination. In analogy to electric charges, disclinations of the
	same sign repel and unlike signs attract according to the continuum elasticity
	theory~\cite{nelson2002defects}. Topological defect is an important entity and
	a useful concept for understanding a series of elastic and plastic behaviors
	of crystals~\cite{Bausch2003e,bowick2009two,Yao2013a,yao2019command}.

	In previous work, we analyzed the static equilibrium distribution of the
	long-range repulsive particles on the disk from the unique perspective of
	topological defects~\cite{Yao2013a}. Specifically, we focused on the
	total topological charge $Q$:
	\begin{eqnarray}
		Q \equiv \sum_{r_i<r'} q_i,
	\end{eqnarray}
	where $r_i$ is the distance of the particle $i$ carrying topological charge $q_i$ to the center of the disk. The sum is over all of the topological charges within the circular
	domain of radius $r'$. $r'$ is smaller than the radius of the disk by one 
	lattice spacing. It is found that due to the intrinsic inhomogeneity created
	by the long-range electrostatic force, the sign of the total topological charge is
	negative, implying the existence of a hyperbolic geometric structure
	underlying the inhomogeneity phenomenon of particle
	packings~\cite{Yao2013a}.

	The static disk model was recently extended to the dynamical regime, and the
	temporally-varying particle configurations on the disk were analyzed in terms
	of topological defects~\cite{yao2021fast}. An important observation is that
	the time-averaged total topological charge $\langle Q \rangle$ is still
	negative in the dynamical regime; the value of $\langle Q \rangle$ is obtained
	by averaging over 100 statistically independent instantaneous particle
	configurations after the system reaches equilibrium. In other words, in the
	statistical sense, the characteristically negative sign of the total
	topological charge as found in the static disk model is preserved in the
	complicated dynamical evolution of the particle configuration. Furthermore, it
	is found that the sign of the time-averaged total topological charge $\langle
	Q \rangle$ switches from positive to negative with the increase in the range
	of interaction.

	\begin{figure}[t] 
		\centering 
		\includegraphics[scale=0.22]{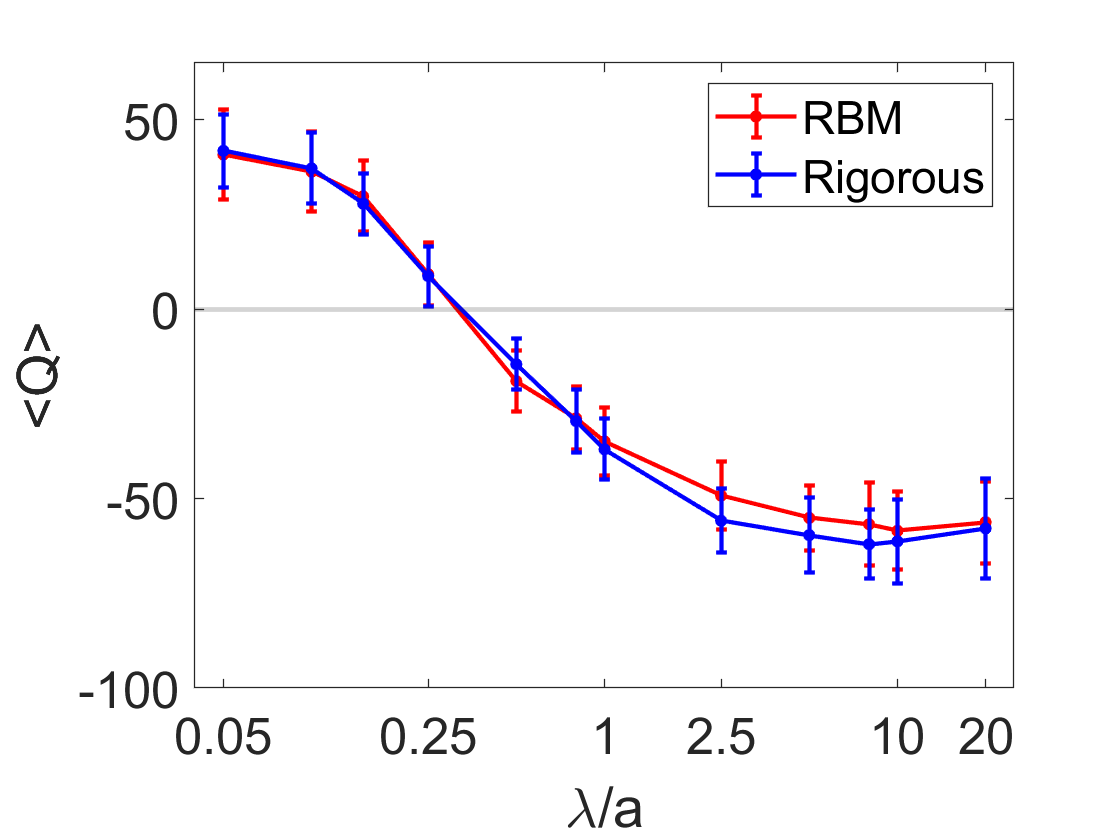}
		\caption{\label{fig:N5000} Statistical regularity revealed from the
			perspective of total topological charge $Q$ in the relatively small system
			of $N=5000$.  The $\langle Q \rangle$-$\lambda$ curves in red and blue
			show the results obtained by the RBM and the rigorous calculations,
			respectively. The statistical analysis of the total topological charge $Q$
			is based on 100 particle configurations of equal time interval
			$0.5\tau_0$ during $t\in [50.5\tau_0, 100\tau_0]$.  The error bars
			indicate the magnitude of the standard deviations of $Q$. The initial
			conditions for both cases are identical.
		} 
	\end{figure}

	\begin{figure*}[t]
		\centering
		\subfigure[$\lambda/a=0.05$]
		{\includegraphics[scale=0.19]{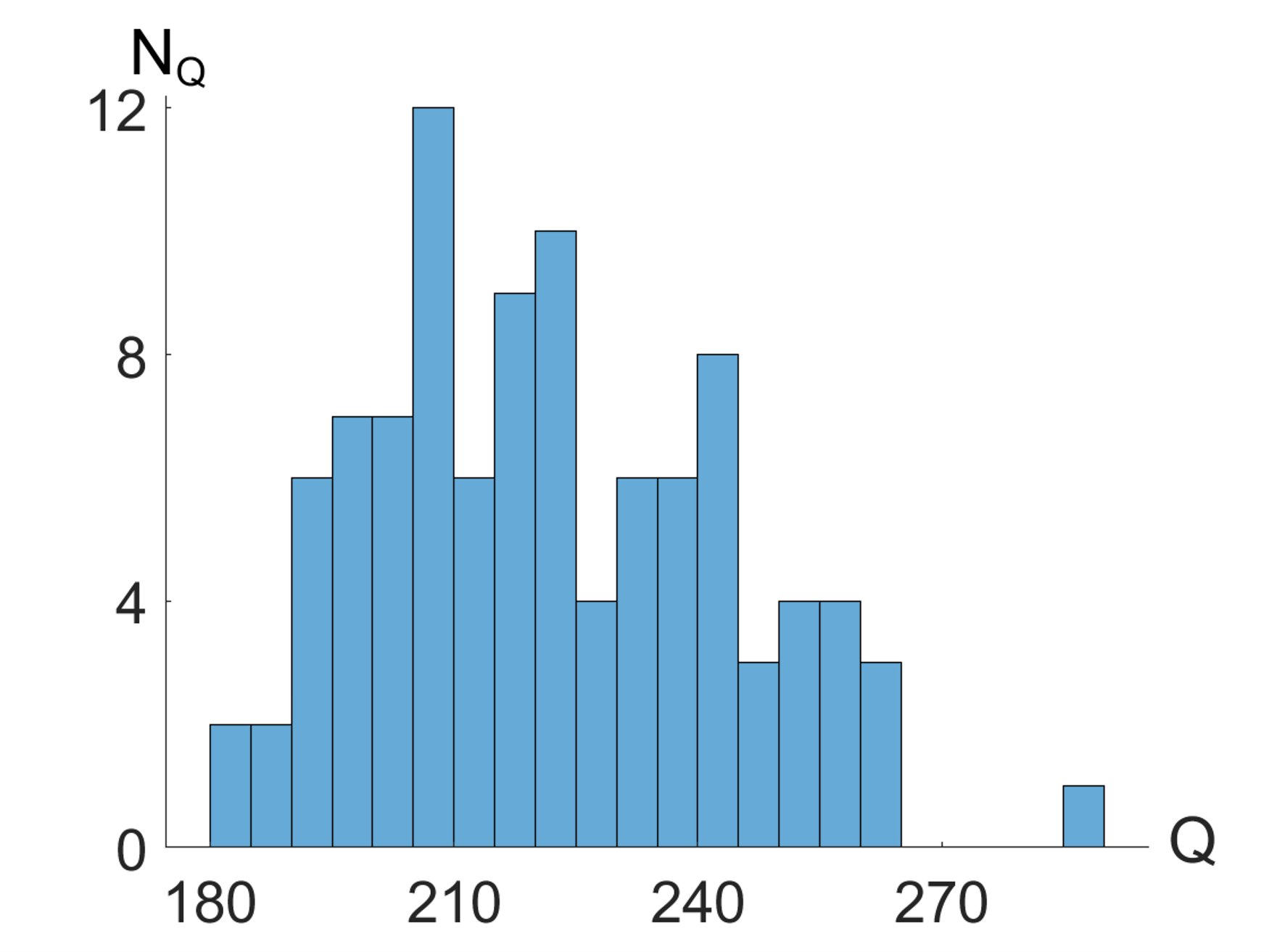} \label{fig:Q lambda his0.05}}
		\subfigure[$\lambda/a=20$]
		{\includegraphics[scale=0.19]{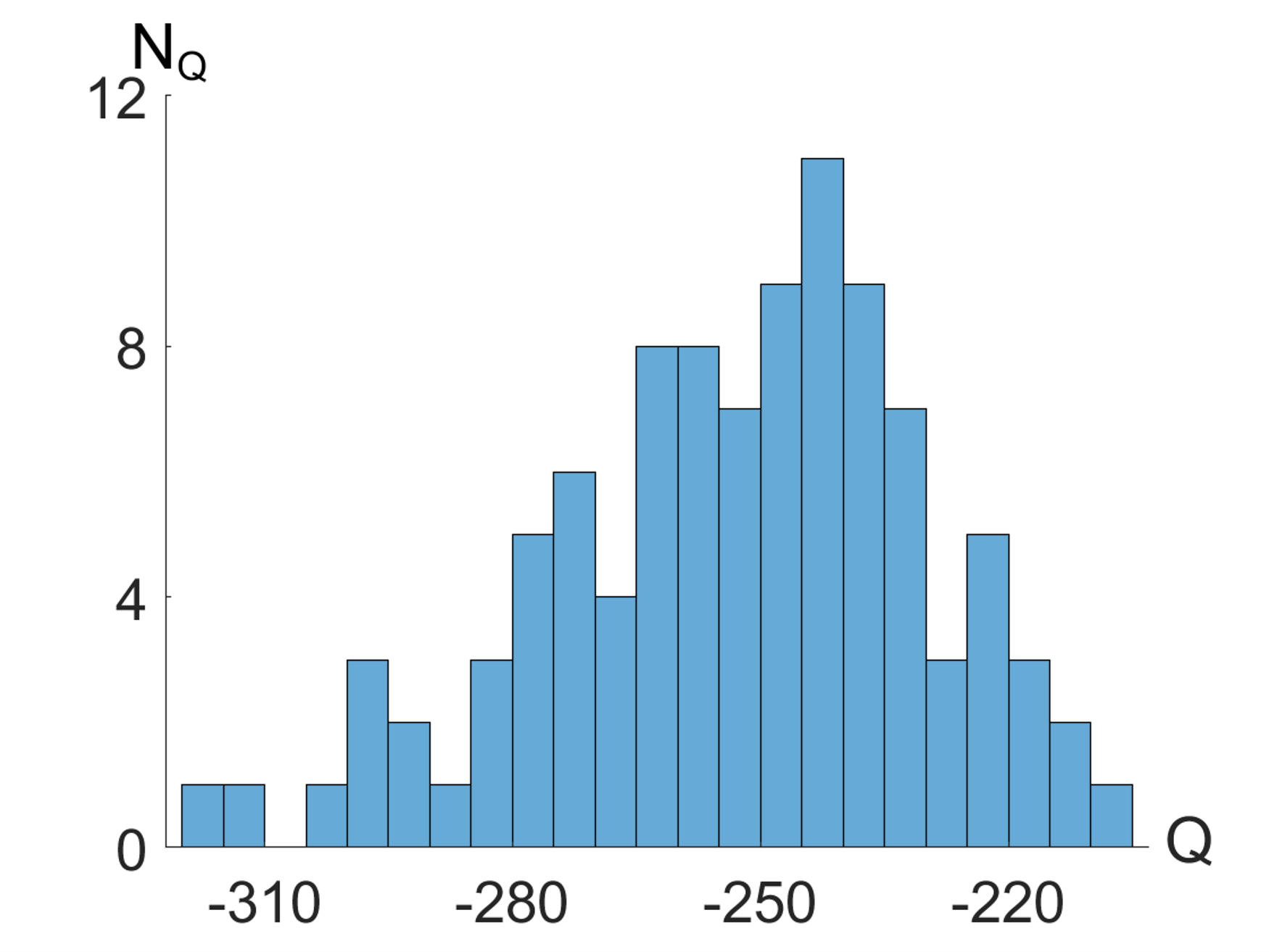} \label{fig:Q lambda his20}}	
		\subfigure[][\label{fig:Q lambda curve}]
		{\includegraphics[scale=0.2]{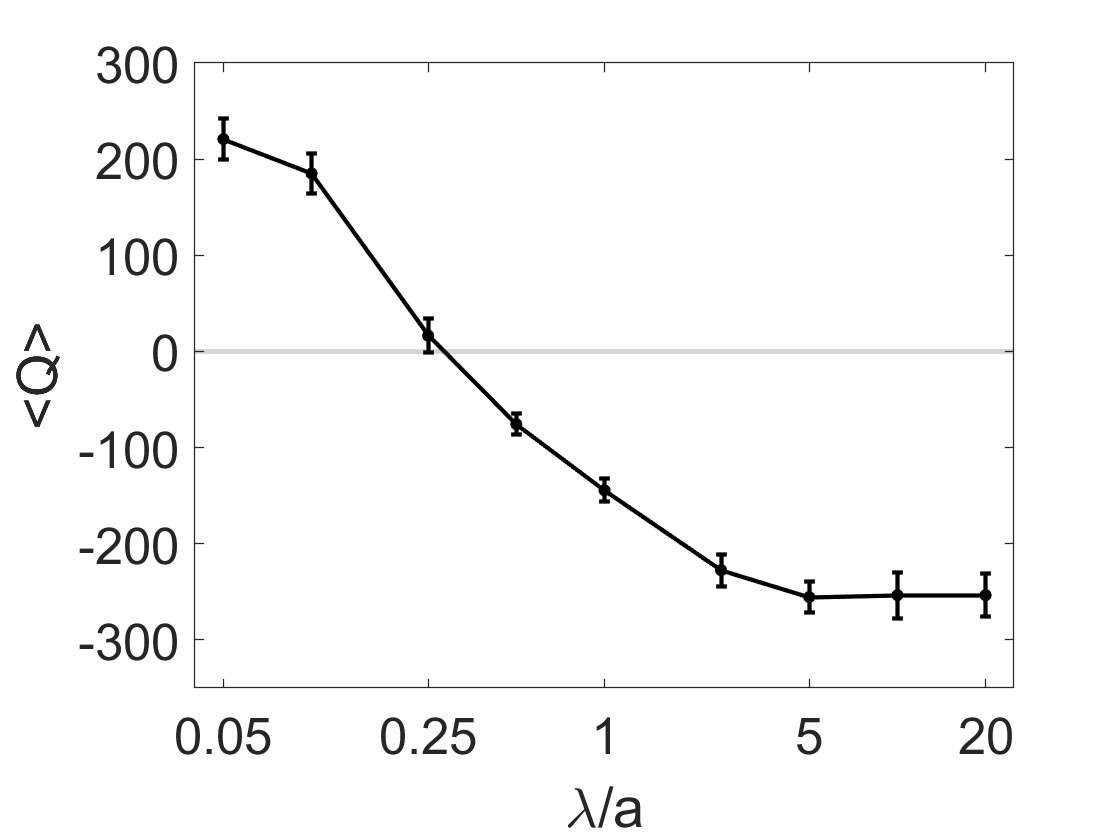}}
		\caption{Statistical regularity revealed from the perspective of total
			topological charge $Q$ in the convoluted dynamical evolution of short- and
			long-range repulsive particles on the disk. (a) and (b) The distributions of
			$Q$ in 100 statistically independent instantaneous particle
			configurations after the system reaches equilibrium. (c) Plot of $\langle
			Q\rangle$ vs the screening length $\lambda$. $\langle Q\rangle$ is the
			time-averaged total topological charge. The error bars indicate the
			magnitude of the standard deviations of $Q$. $N=150\,000$.}
		\label{fig:Q_lambda} 
	\end{figure*}

	However, the revealed statistical regularity in $\langle Q \rangle$ occurs in
	systems consisting of a relatively small number of particles; the maximum
	value of $N$ in the work of Ref.~\cite{yao2021fast} is 5000 as limited by the
	computational time. The boundary effect may be involved. It is
	important to clarify if the statistical regularity in $\langle Q \rangle$ reflects the
	intrinsic effect of the long-range interaction or it is caused by the boundary
	effect.

	To address this question, we shall investigate the regime of
	$\lambda/r_0 \ll 1$, where the boundary effect on the statistical distribution
	of the particles could be reduced. In terms of the number of particles, by
	making use of Eq.~(\ref{ar}),
	\begin{eqnarray}
		\frac{\lambda}{r_0}=\frac{\lambda}{a}\sqrt{\frac{2\pi}{\sqrt{3}N}} \ll 1.
		\label{bc}
	\end{eqnarray}	
	Furthermore, we tune the value of $\lambda/a$ to be much larger than
	unity for exploring the interested regime of long-range interaction. To
	conveniently discuss both conditions of $\lambda/r_0 \ll 1$ and $\lambda/a \gg
	1$, we rewrite Eq.(\ref{bc}) as 
	\begin{equation}
		N=\frac{2\pi}{\sqrt{3}}\left(\frac{\lambda/a}{\lambda/r_0}\right)^2 \gg 1.
		\label{N12}
	\end{equation} 
	Equation~(\ref{N12}) shows that the number of particles shall be
	sufficiently large to simultaneously fulfill the above-mentioned two
	conditions.  For a given value of $\lambda/r_0$, the number of particles
	increases with $\lambda/a$ quadratically. For the example of
	$\lambda/r_0=0.1$, when $\lambda/a = 10$, $N=36\,275$. As the value of
	$\lambda/a$ is increased to 20, $N=145\,103$. By striking a balance of the
	limited computational resources and the required sufficiently large number of
	particles, in simulations we set the value of $N$ as $N=150\,000$, for which
	the disk size is as large as 10 times the screening length even when the
	screening length is up to 20 times the mean lattice spacing. For
	a shorter screening length, the ratio $r_0/\lambda$ is even larger.

  We first test the RBM by applying it to a relatively small system consisting
  of 5000 particles under varying screening length, which has been studied based
  on the rigorous calculations of the interacting forces~\cite{yao2021fast}.  It
  turns out that the $\langle Q \rangle$-$\lambda$ curves based on both the
  RBM and the rigorous calculations are almost identical as shown in
  Fig.~\ref{fig:N5000}. On both $\langle Q \rangle$-$\lambda$ curves, the sign
  of the time-averaged total topological charge $\langle Q \rangle$ turns from
  positive to negative when the value of $\lambda$ is increased to about 0.3;
  the magnitudes of the error bars are also similar.  The invariance of the
  $\langle Q \rangle$-$\lambda$ curve is remarkable, considering that in the
  calculation of the force on each particle only a few randomly picked particles
  instead of all of the other particles are counted. Furthermore, the
  histograms of the total topological charge at typical values of $\lambda$ as
  obtained by the RBM are similar to those obtained by the rigorous calculations of the
  interacting forces.

  Here, we shall point out that due to the approximation of the force in the
  RBM, the resulting microscopic particle trajectories inevitably deviate from
  those obtained by the rigorous calculations of the forces. However, the
  statistical property in question is insensitive to the discrepancy in the
  particle trajectories. Specifically, the agreement of the relevant results
  based on the RBM and the rigorous calculations shows the capability of the RBM
  in calculating the characteristic statistics of the topological charges in the
  temporally-varying particle configurations under both short- and long-range
  repulsions. As such, the RBM serves as a proper tool for exploring the
  underlying statistical regularity in the dynamical disk model. These results
  also suggest the robustness and effectiveness of the quantity $Q$ for
  characterizing the convoluted dynamical evolutions of short- and long-range
  interacting particle systems.

	Now, we employ the RBM to explore the large-$N$ regime. The
	results are summarized in Fig.~\ref{fig:Q_lambda} for $N=150\,000$. We collect
	statistically independent particle configurations in the dynamical evolution
	of the system in equilibrium, and we present the histograms of the total
	topological charge $Q$ for $\lambda/a=0.05$ and $20$ in
	Figs.~\ref{fig:Q lambda his0.05} and \ref{fig:Q_lambda}(b).  It clearly shows that the short-
	and long-range interacting systems could be distinguished by the distribution
	of $Q$. 
	
	Specifically, for the case of $\lambda/a=0.05$ in Fig.~\ref{fig:Q lambda
		his0.05}, the values of $Q$ are uniformly positive in all of the instantaneous
	states. In contrast, for the long-range interacting system of $\lambda/a=20$
	in Fig.~\ref{fig:Q lambda his20}, the values of $Q$ become negative. Here, the
	boundary effect as measured by the ratio $\lambda/r_0$ could be ignored due to
	the large value of $N$. In both cases of $\lambda/a=0.05$ and $20$,
	the value of $r_0$ is larger than $\lambda$ by at least one order of magnitude
	for $N=150\,000$. Therefore, the characteristic distributions of the total
	topological charge $Q$ in Fig.~\ref{fig:Q_lambda} reflect the intrinsic effect
	of the physical interactions on the dynamical organizations of the particles.

	We further inquire how the range of the repulsive force influences the
	distributions of the total topological charge $Q$. To address this question,
	we systematically investigate the dynamical evolution of the particles at
	varying $\lambda$. The quantitative dependence of the time-averaged total
	topological charge $\langle Q\rangle$ on the screening length $\lambda$ is
	shown in Fig.~\ref{fig:Q lambda curve}. We see that the $\langle
	Q\rangle$-curve monotonously decreases with $\lambda$. The turning point for
	the value of $\langle Q\rangle$ switching from positive to negative is located
	at $\lambda_c \approx 0.3$.

	Scrutiny of the $\langle Q\rangle$-curves at varying $N$ show that the value
	of the time-averaged total topological charge $\langle Q\rangle$ increases
	with the increase of $N$. However, the value of $\lambda_c$, where the sign of
	$\langle Q\rangle$ is changed, is insensitive to the variation of $N$. The
	value of $\lambda_c$ is very close to 0.3 as the number of particles is
	increased from $5000$ to $150\,000$. Comparison of Figs.~\ref{fig:N5000} and 3
	shows that the relative fluctuation of $\langle Q\rangle$ (as indicated by the
	relative size of the error bars) decreases with the increase of $N$.

	\begin{figure*}[t]
		\centering
		\includegraphics[scale=0.65]{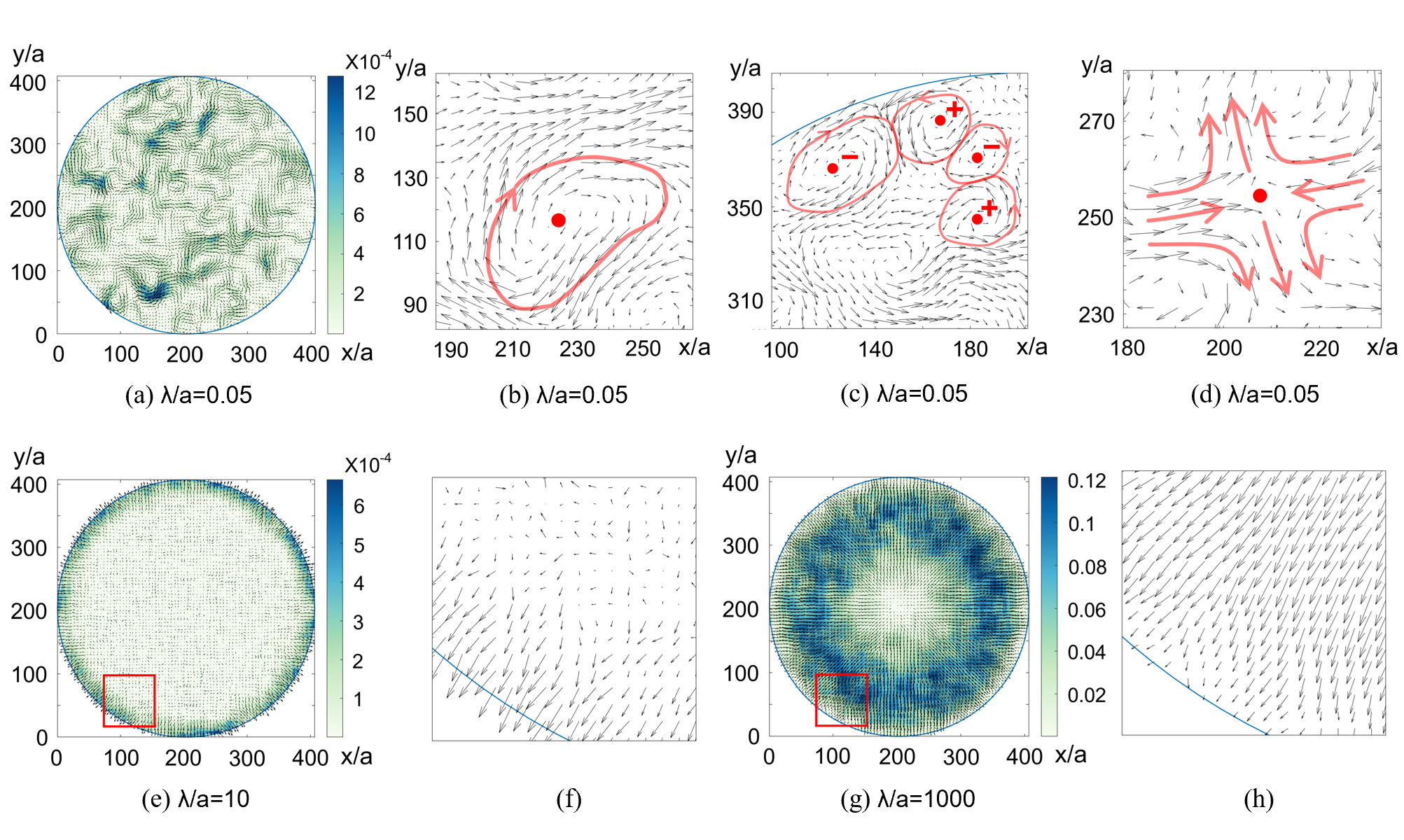}
		\caption{Dynamical organizations of repulsive particles on the disk as
			demonstrated by the spatio-temporally averaged (coarse-grained) velocity
			field. (a), (e), and (g) are the plots of the coarse-grained velocity fields for the cases of $\lambda=0.05$, $\lambda=10$, and $\lambda=1000$, respectively. The color bars besides indicate the magnitude of the kinetic energy. (b)-(d) show the local zoomed-in plots of the velocity field in (a). (f) and (h) show the local zoomed-in plots of the velocity fields in (e) and (g), respectively. The flow patterns under the short-
			and long-range repulsive forces are featured with the vortices of winding
			number $\pm 1$ [(b)-(d)] and the outward radial flow near the boundary [(f)
			and (h)], respectively. The coarse-graining procedure is presented in the
			main text. $N=150\,000$.
		}
		\label{short}
	\end{figure*}

	\begin{figure}[th]
		\centering
		\subfigure[\label{fig:v distribution evolution l0.05}][$\lambda/a=0.05$]
		{\includegraphics[scale=0.27]{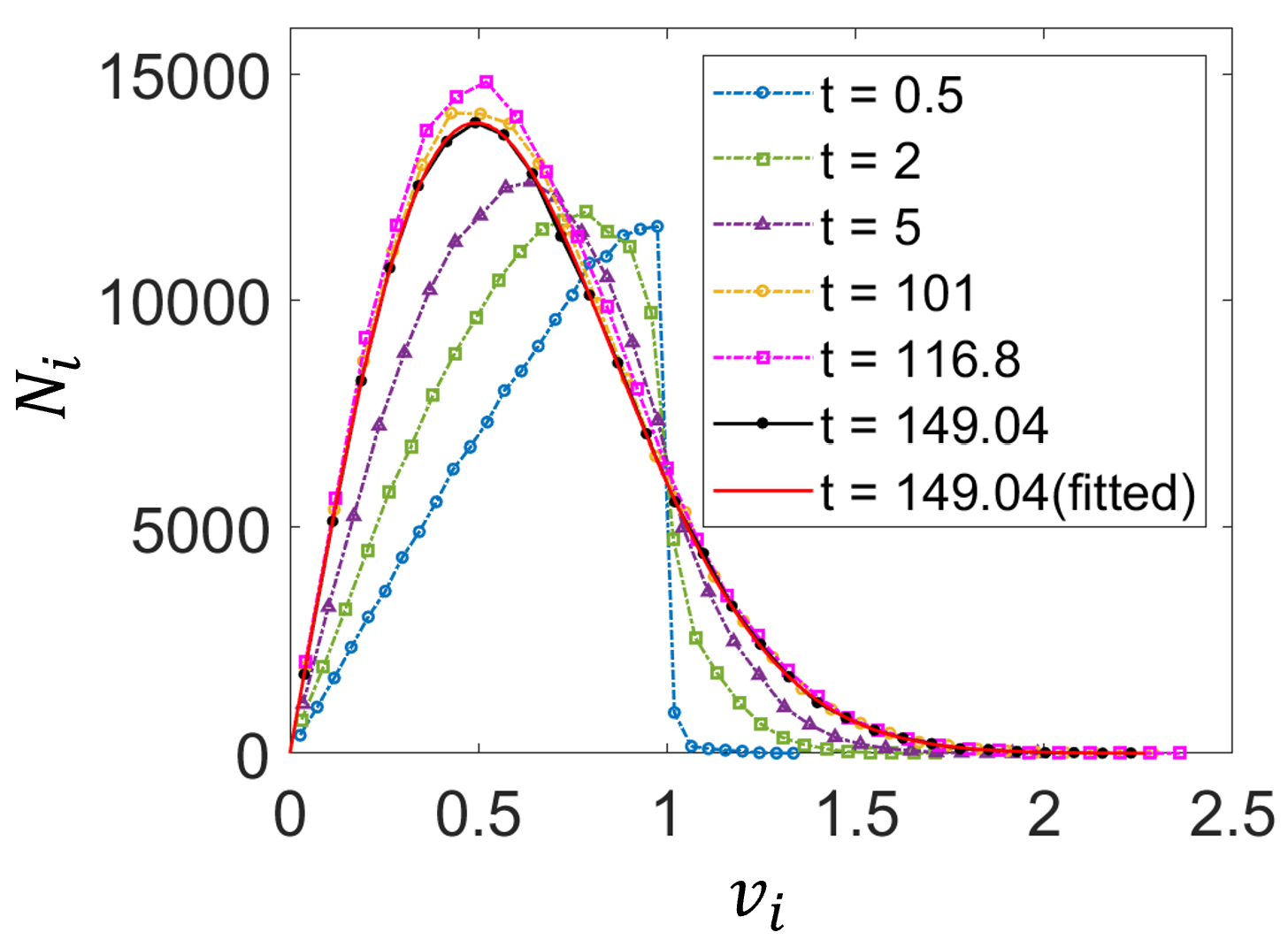}}
		\subfigure[\label{fig:v distribution evolution l10}][$\lambda/a=10$]
		{\includegraphics[scale=0.27]{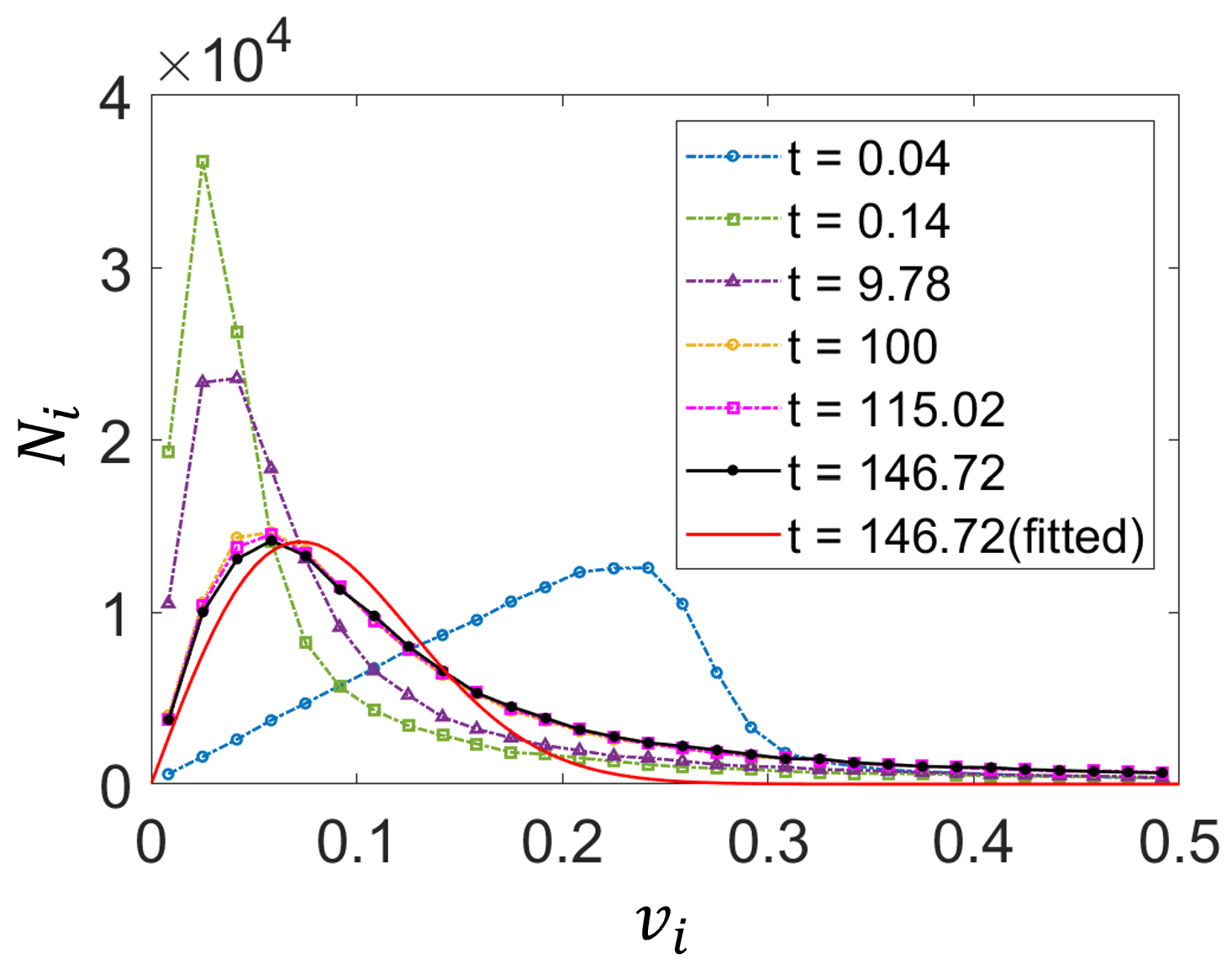}}
		\caption{The relaxation process of particle speed in short- and long-range
			interacting systems. The final distribution curves at $t=149.04\tau_0$ (a) and
			$t=146.72\tau_0$ (b) are fitted by the 2D Maxwell-Boltzmann distribution (solid red
			curves). (b) The deviation of the final distribution curve at
			$t=146.72\tau_0$ (black) and its optimal fitting curve (solid red curve)
			could be attributed to the RBM-induced heating phenomenon.  The plot range
			is for $v \leq 0.5$; about 10\% of the fast particles whose speed exceeds
			$0.5$ are not shown. The values of the fitting parameters: (a)$A=615,476$,
			$\alpha=2.05$, $\delta v=0.0756$; (b)$A=19,179,641$, $\alpha=94.98$, $\delta v
			= 0.0167$. $N=150\,000$.
		}
		\label{fig:speed distribution} 
	\end{figure}

	\subsection{Characteristic coarse-grained velocity field and the relaxation of particle speed}

	The RBM allows one to explore the large-$N$ regime, and to perform a
	coarse-graining procedure for obtaining a smooth velocity field. In this
	subsection, from the perspective of analyzing the coarse-grained velocity
	field, we discuss the dynamical organizations of particles on the disk under
	short- and long-range repulsive forces. Furthermore, we examine the
	relaxation of particle speed in large-$N$ systems of varying $\lambda$, by
	which the validity of the RBM in capturing relaxation dynamics is also
	tested.

  To obtain a smooth coarse-grained velocity field, the circumscribed square of
  the disk is evenly gridded into $n\times n$ square-shaped cells. Each cell is
  associated with a spatio-temporally averaged velocity vector, which is
  obtained by averaging over the velocities of the particles belonging to the specific
  cell and the two layers of surrounding cells for multiple instantaneous
  states of equal time interval $\delta t$. For revealing the
	characteristic behaviors of short- and long-range interacting systems, we
	compare the coarse-grained velocity fields under identical conditions. The
	values of the relevant parameters are listed here: $n=80$, $\delta t =
	2\tau_0$, and $t\in [0.1\tau_0, 60.1\tau_0]$.

	In Figs.~\ref{short}(a), \ref{short}(e) and \ref{short}(g), we show the
	coarse-grained velocity fields of distinct patterns at varying screening
	length; the corresponding zoomed-in plots are also presented. The color bars
	indicate the magnitude of kinetic energy as calculated according to the
	coarse-grained velocity. For the case of $\lambda=0.05$ in
	Figs.~\ref{short}(a)-\ref{short}(d), small-scale dynamical structures are
	developed under the short-range repulsive force. Specifically, we identify
	vortices of winding number $+1$ and $-1$; the red loops and lines are
	introduced for visual convenience.  Figure~\ref{short}(b) shows a deformed
	$+1$ vortex surrounded by a clockwise velocity pattern. According to the
	elasticity theory of vortex, vortices of the same sign repel and unlike signs
	attract~\cite{nelson2002defects,Sethna2006}. Here, in the dynamical regime, we
	also observe a chain of connected $+1$ vortices with alternating clockwise
	and counter-clockwise orientations as shown in Fig.~\ref{short}(c); the
	negative and positive signs represent clockwise and counterclockwise
	directions, respectively.  In Fig.~\ref{short}(d), we show a vortex of winding
	number $-1$. For both $+1$ and $-1$ vortices, the velocity field vanishes at
	the central singular point in the continuum limit. As such, the regions of
	light color in Fig.~\ref{short}(a) correspond to the singular domains of
	vortices. The entire velocity field is essentially composed of vortices of
	winding number $\pm 1$.

	In contrast, the coarse-grained velocity fields at early stage in systems of
	larger screening length are featured with the outward radial flow near the
	boundary, as shown in Figs.~\ref{short}(e)-\ref{short}(h) for the cases of
	$\lambda=10$ and $1000$, respectively. The formation of the radial
	flow pattern could be attributed to the long-range nature of the repulsive
	force; the particles near the boundary are subject to outward radial force
	from interior particles.  The boundary radial flow in the long-range
	repulsive systems in Figs.~\ref{short}(e)-\ref{short}(h) implies that the
	total winding number of the velocity field is $+1$. Note that this flow is
	connected to the phenomenon of density inhomogeneity via the accumulation of
	particles near the boundary in the static equilibrium packings of long-range
	repulsive particles on a disk~\cite{Yao2013a}. Closer examination of the
	zoomed-in velocity fields in Figs.~\ref{short}(f) and \ref{short}(h) shows
	that the region of the radial flow tends to extend towards the center of the
	disk with the increase of the screening length, indicating the enhanced
	alignment of velocity vectors under larger screening length. The ensuing
	velocity fields for both cases of $\lambda=10$ and $1000$ become
	highly irregular.

	Comparison of the distinct velocity fields shaped by the short- and long-range
	forces in Fig.~\ref{short} demonstrates the significant impact of the range of
	interaction on the dynamical pattern. Remarkably, the RBM is able to capture
	the characteristic dynamical structures in the Hamiltonian system consisting
	of over 100000 interacting particles. The validity of the RBM in
	approximating both the deterministic Hamiltonian dynamics and the stochastic dynamics
	has been rigorously discussed from a mathematical
	viewpoint~\cite{jinRandomBatchMethods2020}.  Here, we provide an important
	physical example of a classical particle system that is of wide interest to the
	communities of soft matter physics and statistical
	physics~\cite{cerkaski2015thomson,Mughal2007,soni2018emergent,vortexSilva2019,yao2021fast}.


  We finally briefly discuss the relaxation process of the distribution of
  particle speed in both short- and long-range interacting systems by the RBM.
  According to the canonical ensemble formalism, the equilibrium distributions
  of particle position and momentum are statistically independent, if the
  Hamiltonian can be written as a sum of two terms containing coordinates and
  momenta, respectively. Thus, regardless of the range of 
  interaction, the distribution of the particle speed is expected to converge to the
  Maxwell-Boltzmann distribution for the dynamical disk model consisting of a
  large number of particles.

  The results for the cases of $\lambda/a=0.05$ and $10$ are presented
  in Fig.~\ref{fig:speed distribution}. We see that in both cases, the
  distribution curves indeed tend to converge to the 2D Maxwell-Boltzmann
  distribution (solid red curves):
	\begin{eqnarray}
		f(v) \delta v =Av\exp(-\alpha v^2) \delta v,
	\end{eqnarray}
	where $A$ and $\alpha$ are fitting parameters.  For the case of
	$\lambda/a=0.05$ in Fig.~\ref{fig:speed distribution}(a), the final
	distribution curve at $t=149.04\tau_0$ (black) agrees well with the 2D
	Maxwell-Boltzmann distribution.  For the case of $\lambda/a=10$ in
  Fig.~\ref{fig:speed distribution}(b), we notice a slight deviation of the
  final distribution curve at $t=146.72\tau_0$ (black) and its optimal fitting
  curve (red). This deviation could be attributed to the RBM-induced heating
  phenomenon and the procedure of cooling in simulations, which is implemented
  for the sake of the conservation of total energy. Note that the heating effect
  caused by the RBM is even more pronounced for the long-range interacting
  system; more information is provided in Appendix B.

  From Fig.~\ref{fig:speed distribution}, we also see the distinct kinetic
  pathways of the relaxation process for the short- and long-range interacting
  systems. For the system of $\lambda/a=0.05$ in Fig.~\ref{fig:speed
  distribution}(a), the peak of the distribution curve keeps moving leftward
  to approach the location of the peak in the equilibrium curve.  This could be
  understood from the perspective of energy transfer.  Specifically, 
  kinetic energy is converted into potential energy in the relaxation of the
  short-range interacting system (see Appendix B). In contrast, for the system
  of $\lambda/a=10$ in Fig.~\ref{fig:speed distribution}(b), we observe the
  initial leftward and the subsequent rightward movement of the peak of the
  distribution curve; the rightward movement after $t \approx 2 \tau_0$
  corresponds to the conversion of potential energy to kinetic energy.
  To conclude, the relaxation process, as reflected in the movement of the peak
  of the speed distribution curve, is classified into two categories: the
  ``overdamped" and the ``underdamped" relaxation dynamics for the short- and
  long-range interacting systems, respectively. These observations imply that
  the short- and long-range interacting systems admit distinct microscopic
  dynamical modes for the relaxation of the speed distribution.

	\section{Conclusion}

	In summary, we investigate the statistical and dynamical physics of the
	dynamical disk model consisting of a large number of repulsive particles. The
	recently developed powerful computational tool of the random batch method, which
	is specifically designed for reducing the computational complexity of
	long-range interacting particle systems down to the order of
	$O(N)$~\cite{jin2022RBM_second_order,jin2020random}, 
	allows us to explore the
	large-$N$ regime of interest. To seek the order underlying the
	convoluted dynamical evolution of the large number of particles, we resort to
	the concept of a topological defect, and we reveal the intrinsic statistical regularity of
	topological charges that is otherwise unattainable by the continuum analysis
	of particle density. We further show the distinct dynamical organizations of the
	particles under short- and long-range repulsive forces.  This work extends the
	disk model from the
	static~\cite{Mughal2007,miguel2011laminar,Yao2013a,
		cerkaski2015thomson,soni2018emergent,silva2020formation,PhysRevE.104.034614}
	to the dynamical regime. We highlight the crucial role of topological
	defects in elucidating the intrinsic statistical order underlying the convoluted
	dynamical organizations of interacting particles. This work also demonstrates the promising potential of the
  random batch method for exploring fundamental scientific questions arising in
  a variety of long-range interacting particle systems in soft matter physics
  and other relevant
  fields~\cite{Bausch2003e,bowick2009two,Campa2014,soni2018emergent,yao2019command}.

	\section{Acknowledgements}
	
	This work was supported by the National Natural Science Foundation of China
	(Grant No. BC4190050).

	\section*{Appendix A: Technical details of simulations}
	
	In this appendix, we present more information about the technical details of
	the boundary condition and the initial condition, and the treatment of abnormally fast
	particles.
	
	In the model, we employ the reflecting boundary to confine the point
	particles in the disk. For a particle that is about to collide with the boundary
	in a simulation step, we first identify the crossing point of the straight
	particle trajectory and the circular boundary, and then we let the particle make
	a mirror reflection with respect to the cross point.  The final speed of the
	particle is the same as the initial incident speed prior to the reflection.

	The initial state of the system is prepared by specifying a random velocity
	$\vec{v}_{\mathrm{ini}}$ and a random position $\vec{r}_{\mathrm{ini}}$ to each particle. The
	magnitude and the direction of the velocity $\vec{v}_{\mathrm{ini}}$ are uniform random
	variables. $v_{\mathrm{ini}}\in (0,1)$ and $\theta \in (0,2\pi)$, where $\theta$ is the
	angle of the velocity vector $\vec{v}_{\mathrm{ini}}$ with respect to some reference
	direction. $\vec{r}_{\mathrm{ini}}$ is determined by the standard procedure of random
	disk packing~\cite{lubachevsky1990geometric}. The radius of the disk is set to
	be $0.3a$, where $a$ is the mean distance of nearest particles. The values of
	the initial positions of the point particles in our model are given by the
	centers of the disks. The random disk packing prevents the aggregation of
	particles, which may result in a large force in a particle.

	In simulations of long-range interacting systems, it is found that the speed of
	a small fraction of the particles (on the order of 10 in the system containing
	150,000 particles)	may become abnormally large; such particles cover a
	long distance over several $a$ (mean distance of nearest particles) in a
	single simulation step.  One may set the time step to be sufficiently fine to
	solve this problem, but such an operation is highly time consuming.  In
	simulations, the speed of these particles is set to be zero. This operation
	seems to have no effect on the statistical properties of interest by a comparison with
	the rigorous simulations of smaller-$N$ systems.

	\section*{Appendix B: Conservation of total energy by cooling procedure}

	\begin{figure}[t]
		\centering
		\subfigure[\label{fig:E_variation_0.05}][$\lambda/a=0.05$]
		{\includegraphics[scale=0.24]{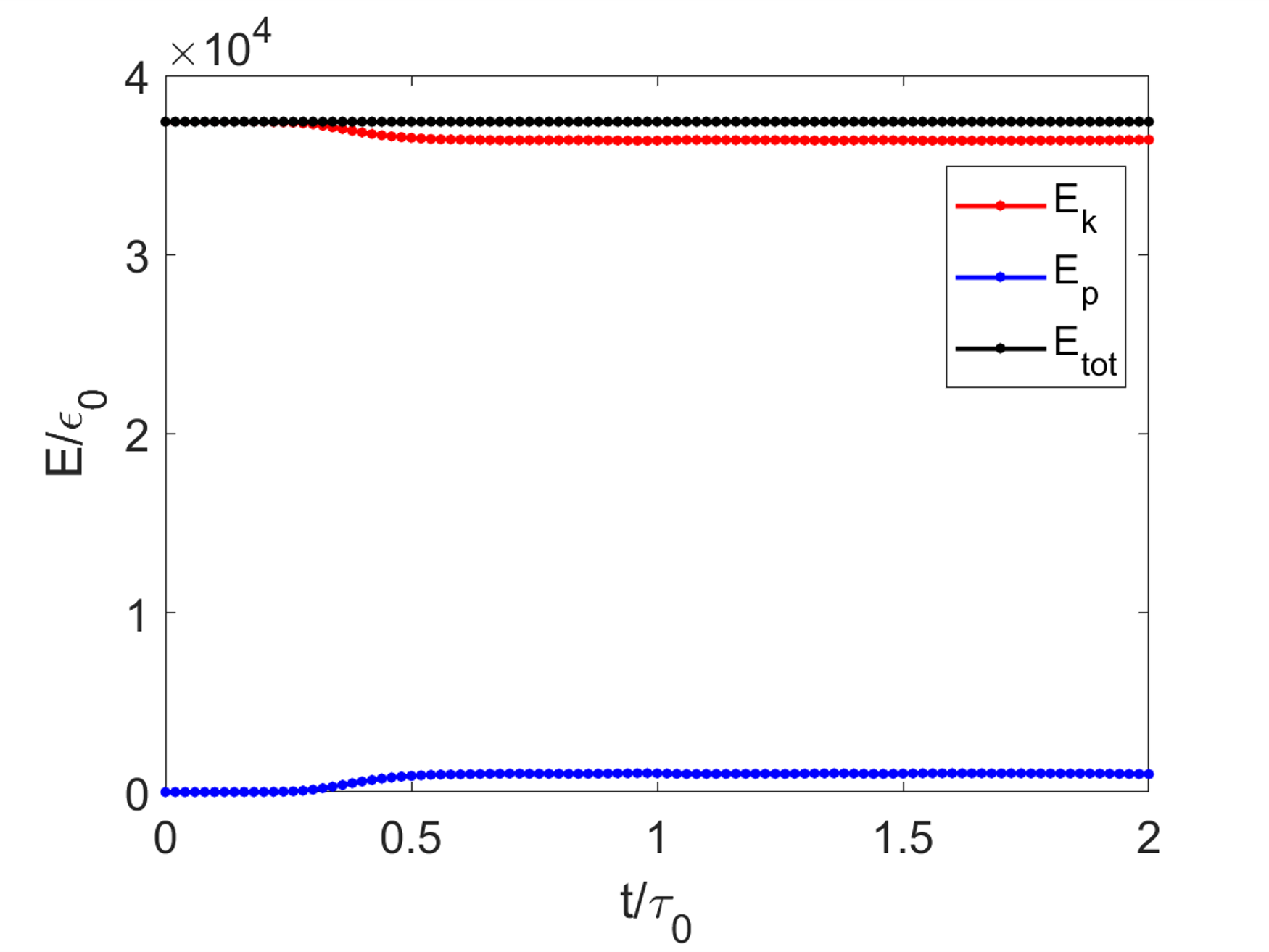}}
		\subfigure[\label{fig:E_variation_10}][$\lambda/a=10$]
		{\includegraphics[scale=0.29]{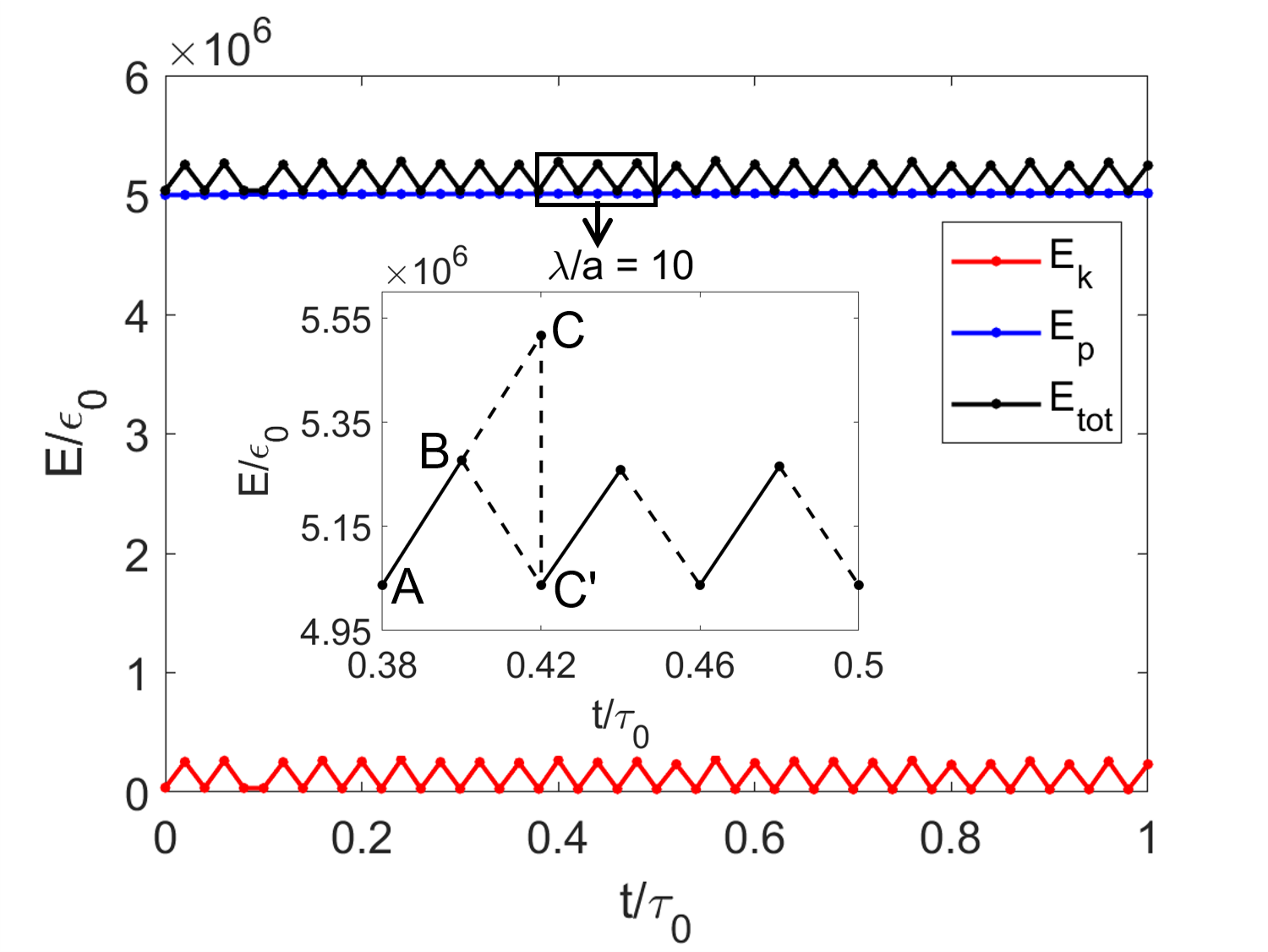}}	
		\caption{The variation of kinetic and potential energies in the dynamical
			evolution of short- and long-range repulsive particles on the disk. The
			inset in (b) illustrates the cooling procedure employed in simulations for
			the sake of the conservation of the total energy. $N=150\,000$.}
		\label{fig:E} 	
	\end{figure}

	The RBM allows one to explore the large-$N$ regime. However, it has been
	reported that the RBM may lead to the heating effect in molecular-dynamics
	simulations~\cite{jin2022RBM_second_order}. In our simulations, it is also found that
	the kinetic energy (temperature) increases monotonously in time. The heating
	effect could be tracked down to the partial and random selection of the 
	particles in the calculation of the total force on a particle.  Consequently,
  the RBM brings in temporally-varying random forces on the particles, which
  causes the monotonous increase of the kinetic energy (temperature). Here, we
  shall point out the robust statistical properties of topological charges
  in the presence of the thermal (random) motion of the particles caused by the RBM
  algorithm. A plausible explanation is that the magnitude of the thermal
  (random) motion shall exceed some critical value (comparable with the lattice
  spacing) to change the coordination number of a particle and thus to generate
  topological charge.

  We employ the following cooling procedure to ensure the conservation of the
  total energy within some tolerance in the dynamical evolution of the system.
  Specifically, the energy of the system is checked periodically, and if the
  amount of the increased energy exceeds some threshold value (5\% of the total
  energy $E^{(0)}_{\mathrm{tot}}$ in the initial state), all of the velocity vectors are
  rescaled by a common factor $\sqrt{(E^{(0)}_{\mathrm{tot}}-E_p)/E_k}$ to ensure that
  the total energy is corrected to its initial value. $E_k$ and $E_p$ are the
  kinetic energy and the potential energy at the time when the system energy is
  checked.

	The temporally-varying energy curves in both short- and long-range interacting
	systems are shown in Fig.~\ref{fig:E}. For the case of $\lambda/a=0.05$ in
	Fig.~\ref{fig:E}(a), the total energy is well conserved during the simulation
	without resorting to the cooling procedure. This could be attributed to the
	rigorous calculations of the interacting forces between adjacent particles in the
	RBM. For the case of $\lambda/a=10$ in Fig.~\ref{fig:E}(b), the cooling
	procedure is triggered when the total energy increases to be above 5\% of its initial
	value, as indicated by the dashed vertical line from C to C' in the zoomed-in
	inset. Note that at the time of the checkpoint, the total
	energy of the system may have exceeded 5\% of its initial value; for example, at
	the checkpoint C, the value of the total energy is about $9\%$ of its initial
	value. In the data analysis, we select instantaneous states whose total energy
	is approximately constant; the variation of the total energy is within 5\% of
	its initial value, as indicated by the solid black lines in
	Fig.~\ref{fig:E}(b).


\end{document}